\documentclass[preprint,nofootinbib]{revtex4-1}
 \usepackage{graphicx,psfrag,epsfig}
\begin{document}
\begin{flushright}
UMD-PP-012-002, MIFPA-12-05\\
June 2012
\end{flushright}
\title {\LARGE $\theta_{13}$ and Proton Lifetime in a Minimal  $SO(10)\times S_4$ Model of Flavor}
\author {\bf   P. S. Bhupal Dev$^1$,  Bhaskar Dutta$^2$, R. N.
Mohapatra$^1$ and Matthew Severson$^1$\\}
\affiliation{ $^1$ Maryland Center for Fundamental Physics and
Department of Physics, University of Maryland, College Park, MD
20742, USA}
\affiliation{$^2$ Department of Physics \& Astronomy, Mitchell Institute for 
Fundamental Physics, Texas A \& M University, College Station, TX 77843-4242, USA}
\begin{abstract}
In a recent paper, a minimal supersymmetric (SUSY) $SO(10)\times S_4$ based unified model of  flavor for  quarks and leptons was proposed with two {\bf 10} and one {\bf 126} contributing to fermion masses. An important aspect of this model is that Yukawa couplings  emerge  dynamically from minimization of the flavon 
potential, thereby reducing the number of parameters considerably. We make a detailed numerical analysis of this model for fermion mixings including SUSY threshold effects at the TeV scale and type-I corrections to 
a type-II dominant seesaw for neutrino masses. This is a single-step breaking model with SUSY $SO(10)$ broken at the Grand Unified Theory (GUT)-scale of 
$2\times 10^{16}$ GeV to the Minimal Supersymmetric Standard Model (MSSM). 
The minimal model has only 11 parameters, and therefore, the charged fermion fits 
predict the masses (up to an overall scale) and mixings in the neutrino sector. 
We present correlations for the different predictions in the neutrino mixing 
parameters. The recent experimental ``large'' $\theta_{13}$ value of $\sim 9^
\circ$ can be obtained by a simple extension of the minimal model.  
We also find that  proton decay 
mode $p\to K^+\bar{\nu}_\mu$ has a partial lifetime of $\sim 10^{34}$ yrs, which is within reach of the next round of planned proton decay searches. 
The successful fit for fermion masses requires the Higgs mass to be below 129 GeV in this model. If the Higgs mass lies between 120-128 GeV, as suggested by the recent LHC data, we find a lower limit on the light stop mass of 755 (211) GeV for $\mu>0~(<0)$.   
\end{abstract}
\maketitle
\section{Introduction}
The measurement of neutrino masses and mixings during the past decade has provided the first evidence for physics beyond the standard model (SM). It has also raised the possibility that this new knowledge may  help to unravel  the physics of flavor and also possibly unlock the mystery behind the origin of matter in the universe. An important recent experimental finding in this area has been the announcements by the T2K~\cite{T2K}, MINOS~\cite{MINOS} and 
Double CHOOZ~\cite{dchooz}, and most recently by the Daya Bay and RENO~\cite{DayaBay,reno}   
experiments that one of the hitherto unknown neutrino  mixing angles, namely $\theta_{13}$, is not only non-zero but ``large''. 
Some hints for a large $\theta_{13}$ have also been suggested by a global analysis of the existing oscillation data~\cite{lisinew,lisi}. A non-zero $\theta_{13}$ 
has profound implications for our understanding of the physics of neutrino mass, and it is therefore timely to search for the prediction of 
$\theta_{13}$ in various models~\cite{models}. 

A key question before theorists now is: what is the big picture of flavor where known quark and lepton masses and mixings fit together? A framework that suggests itself is grand unified theories (GUT), where all matter and all forces 
(except gravity) unify at high energy scale. Since single-step coupling unification in these theories requires supersymmetry (SUSY), we will focus on SUSY-GUTs for approaching the flavor problem and use the seesaw mechanism~\cite{seesaw} to understand the small neutrino masses. A possible advantage of this framework is that matter unification is expected to reduce the number of free parameters that determine fermion masses and mixings from 31 parameters in the seesaw extended standard model, so that one can make predictions to make the model testable. The minimal scenario which is predictive for neutrinos is a renormalizable supersymmetric $SO(10)$ model with {\bf 10} and {\bf 126} Higgs fields contributing to fermion masses~\cite{babu}. This model embodies both the type-I and type-II~\cite{type2} seesaw contributions to neutrino masses and has been analyzed in many papers~\cite{vissani,goh,many}. Fitting charged fermion masses in these models leads to predictions for lepton mixing angles and will be tested with higher precision measurement of these angles and the neutrino mass hierarchy. Detailed analysis of the superpotential and symmetry breaking for these models have been 
carried out in Ref.~\cite{aulakh}.

The next step with these models is to see if Yukawa couplings can be predicted as consequences of higher-scale symmetries. There have been several such approaches~\cite{alta,DMM,king,king1} which adopt the point of view that Yukawa couplings are dynamical fields, namely the flavon fields, whose vacuum expectation values (vevs) determine the observed Yukawa couplings. If these vevs could be the result of the minimization of simple superpotentials, that would reduce the number of parameters and would indeed be a step further in the search for an understanding of flavor. In a recent paper~\cite{DMM}, such a model was presented where the GUT scale $SO(10)$ theory was extended to include three flavon fields with an $SO(10)\times S_4$ symmetry, where the flavon fields, as well as the three matter families, form a three dimensional irreducible representation of $S_4$.

It was shown in~\cite{DMM} that the ground state of the $SO(10)\times S_4$ theory has only 11 parameters describing all flavor, i.e. quark and charged lepton masses and quark mixings as well as neutrino masses and lepton mixings. Using pure type II seesaw contribution, it was argued that the model is in qualitative agreement with observations.  In this paper we present a detailed numerical analysis of this model and its predictions in the neutrino sector as well as 
for proton decay. 
We find that once we include the SUSY threshold corrections to quark masses and a small type-I contribution to neutrino masses, all existing data in the quark and charged lepton sector can be fitted to a good accuracy; furthermore, the model predicts the solar to atmospheric mass ratio $\Delta m^2_\odot / \Delta m^2_{\rm atm}$ and the solar and atmospheric mixing angles in agreement with the current neutrino oscillation data. 
However, with the flavon vacuum alignment as in Ref.~\cite{DMM}, we find that 
the fermion fit predicts a reactor mixing angle $\theta_{13} \sim 5^\circ$ 
which is more than $3\sigma$ below the current experimental 
value~\cite{DayaBay,reno}. We find that a larger value of $\theta_{13}$, 
consistent with the Daya Bay and RENO results can be obtained by a slightly 
different flavon vacuum alignment which results due to an additional term in the superpotential allowed by the $S_4$ symmetry.  
We will demonstrate the determination of these features, and also note the correlation between the different parameters, which can make it easier to rule out the model. 

We also find that in order to get the desired threshold corrections to the $b$-quark mass to fit observations, we need a large negative 
$\mu$ and/or $A$-terms in the model. We discuss predictions for Higgs mass for this choice of MSSM parameters, and find that the Higgs mass should be below 
129 GeV in order to satisfy the constraints from the fermion sector fit. We also obtain lower limits on the squark masses from the same 
constraints. In particular, we note that if the Higgs mass is discovered between 120-128 GeV, the light stop should be heavier than 755 (211) GeV for 
$\mu>0$ ($\mu<0$) in this model. 
These features could be used to test the model at the LHC.   

Finally, we discuss proton lifetime predictions in this model. We find that in order to suppress the $RRRR$ contributions to proton lifetime, we need to work in the low $\tan\beta$ regime, which is the reason we need a large $A$-term for the threshold 
correction, as noted above.  We find that the $B$-violating  $LLLL$ - terms contribute dominantly to $p\to K^+\bar{\nu}_\mu$ decay mode for which we find a partial lifetime of $\sim 10^{34}$ yrs. We emphasize that proton lifetime predictions as well as the predictions for neutrino mixings can be used to test the model.

The paper is organized as follows: in Section II, we present the essential points of the minimal $SO(10)\times S_4$ model; in Section III, we discuss the fermion mass fits and the predictions for the neutrino sector; in Section IV, we present a slightly different vacuum alignment which predicts a large $\theta_{13}$ 
while being consistent with the rest of the fermion sector. In Section V, we discuss the SUSY threshold correction required to fit quark masses 
and its implications for gluino, stop and the Higgs masses in the model. In Section VI, we discuss our predictions for proton lifetime and in Section VII, we summarize our results.

\section{ Details of the Model}
The class of $SO(10)$ models we are interested in here have two {\bf 10} Higgs fields (denoted by $H,H'$) and one $\overline{\bf 126}$ (accompanied by 
{\bf 126}, denoted by $\Delta+\bar{\Delta}$). The $SO(10)$-invariant Yukawa couplings of the model are given by:
\begin{eqnarray}
{\cal L}_Y~=~h\psi\psi H~+~h'\psi\psi H'~+~f\psi\psi\bar{\Delta}
\end{eqnarray}
where $\psi$'s denote the {\bf 16} dimensional spinors of $SO(10)$, which contain all the matter fields of each generation; there are, of course, three such fields, though we have suppressed the generation indices. The Yukawa couplings are therefore $3\times 3$ matrices in generation space.

The effective Yukawa couplings $f, h, h'$ are assumed to have
descended from a higher scale theory which has an $S_4$ symmetry
broken by flavon fields $\phi_i$ determined by the minimum of the
flavon potential. The alignment of the vacuum expectation value (vev) of the 
flavon fields as given in
Ref.~\cite{DMM} are:
\begin{eqnarray}
	\phi_1 = \left(\begin{array}{c} 0 \\ 0 \\ 1 
	\end{array}\right),~ 
	\phi_2 = \left(\begin{array}{c} 0 \\ -1 \\ 1
	\end{array}\right),~ 
	\phi_3 = \left(\begin{array}{c} 1 \\ 1 \\ 1
	\end{array}\right).
	\label{eq:flavev}
\end{eqnarray}
As noted in Ref.~\cite{DMM}, in order to get the desired Yukawa couplings naturally from the high scale theory, we supplement the $S_4$ symmetry group by an 
$Z_n$ group, and the corresponding effective superpotential is given by
\begin{eqnarray}
	W = (\phi_1\psi)(\phi_1\psi)H+(\phi_2\psi)(\phi_2\psi)\bar{\Delta}+
	(\phi_3\psi\psi)\bar{\Delta}+(\phi_2\psi\psi)H'
	\label{eq:supmin}
\end{eqnarray}
where the brackets stand for the $S_4$ singlet contraction of flavor 
index~\cite{hagedorn}. 

The fermion mass matrices are derived from the Yukawa interaction as follows: after GUT symmetry breaking, two linear combinations of the SM doublets remain 
light, denoted by $H_u$ and $H_d$. Typically, $H_{u(d)}=\sum_\alpha U_{u(d)\alpha} H_{u(d)\alpha}$, where $H_{u(d)\alpha}$ are the up(down)-type SM doublets in the $(H, H', \bar{\Delta})$. The effective $H_u$ coupling at the MSSM scale is then given by $QH_uu^c(hU_{uH}+fU_{u\Delta}+h' U_{uH'})$, and similarly for 
$H_d$. The fermion mass matrices can be written in terms of these couplings as
\begin{eqnarray}
{\cal M}_u &=& \bar{h}+r_2 \bar{f}+r_3\bar{h}^\prime \nonumber \\ 
{\cal M}_d &=& \frac{r_1}{\tan\beta}
(\bar{h}+\bar{f}+\bar{h}^\prime) \nonumber \\
{\cal M}_\ell &=& \frac{r_1}{\tan\beta}
(\bar{h}-3\bar{f}+\bar{h}^\prime) \nonumber \\
{\cal M}_{\nu_D} &=& \bar{h} - 3 r_2 \bar{f} + r_3 \bar{h}^\prime,
\label{eq:mass}
\end{eqnarray}
where we have absorbed the mixings $(U_{u\alpha},U_{d\alpha})$ and the 
vevs $\langle H_{u,d} \rangle = \kappa_{u,d}$, with 
$\tan\beta=\kappa_u/\kappa_d$, to re-define the Yukawa coupling 
matrices as follows:  
\begin{eqnarray}
\bar{h} = \kappa_u U_{uH} h, ~ ~ 
\bar{f} = \frac{\kappa_u U_{d\Delta}}{r_1}f, ~ ~
\bar{h^\prime} = \frac{\kappa_u U_{dH'}}{r_1}h^\prime,
\label{eq:couplings}
\end{eqnarray}
with the ratios 
\begin{eqnarray}
r_1 = \frac{U_{dH}}{U_{uH}},~ ~ 
r_2 = r_1\frac{U_{u\Delta}}{U_{d\Delta}},~ ~
r_3 = r_1\frac{U_{uH'}}{U_{dH'}}.
\label{eq:r}
\end{eqnarray}

These effective coupling matrices determined by the flavon sector 
are of the form  
\begin{eqnarray}
	\bar{h} &=& \left(\begin{array}{ccc} 
0 & 0 & 0 \\ 0 & 0 & 0 \\0 & 0 & M\end{array}\right), \\ 
\bar{f} &=& \left(\begin{array}{ccc}0 & m_1 & m_1 \\
m_1 & m_0 & m_1-m_0\\
m_1 & m_1-m_0 & m_0
\end{array}\right), \label{eq:f} \\
\bar{h}' &=& \left(\begin{array}{ccc} 
0 & \delta & -\delta \\ \delta & 0 & 0 \\-\delta & 0 & 0
\end{array}\right) \label{eq:hp}.
\end{eqnarray}
It was shown in Ref.~\cite{DMM} that $S_4$ symmetry constrains $\bar{h}$ to 
have the above rank-one form. The parameters $m_0, m_1, \delta$ are chosen to be complex, giving a total of 10 parameters in the charged-fermion sector.

The neutrino mass matrix is, in general, given by a combination of the type-I 
and II seesaw mechanism:
\begin{equation}
	{\cal M}_\nu~=~v_L f - {\cal M}_{\nu_D}\left(v_R f\right)^{-1}\left(
	{\cal M}_{\nu_D}\right)^T,
\label{eq:neu}
\end{equation}
where $v_{L,R}$ are the vevs of the SM triplet Higgs fields 
$\Delta_{L,R}$ in {\bf 126}. 
If we assume type-II dominance, imposed by the ratio of $v_L$ and $v_R$, and 
the magnitude of the coupling $f$, the neutrino mass matrix ${\cal M}_\nu$ 
takes an approximate tri-bimaximal (TBM) ~\cite{tbm} 
form via the form of $f$ given in Eq.~(\ref{eq:f}), which is diagonalized by
\begin{eqnarray}
	V_{\rm TBM} = \left(\begin{array}{ccc}
		\sqrt{\frac{2}{3}} & \sqrt{\frac{1}{3}} & 0\\
		-\sqrt{\frac{1}{6}} & \sqrt{\frac{1}{3}} & -\sqrt{\frac{1}{2}}\\
		-\sqrt{\frac{1}{6}} & \sqrt{\frac{1}{3}} & \sqrt{\frac{1}{2}}
	\end{array}\right)
	\label{eq:tbm}
\end{eqnarray}
Note that the full neutrino mixing matrix, $U_{PMNS}=V^\dagger_{\ell}V_\nu$, 
where $V_{\ell}$ and $V_{\nu}$ are the unitary matrices that diagonalize the 
charged lepton and neutrino mass matrices, respectively.  Hence, we will 
necessarily have corrections to the TBM mixing given by Eq.~(\ref{eq:tbm}) 
coming from the charged lepton mass matrix as well as the type-I contribution.  
Note further that since the $f$ matrix also contributes to the quark and charged 
lepton masses, neutrino masses and quark masses are connected, thus 
making the model predictive.
\section{Predictions in the Neutrino Sector}
We diagonalize the mass matrices given by Eqs.~(\ref{eq:mass}) and use 
the {\tt Minuit2} tool library~\cite{minuit} to 
minimize the sum of chi-squares for the mass eigenvalues and CKM mixing in the 
charged-fermion sector as well as the mass-squared differences 
$\Delta m_\odot^2$ and $\Delta m_{\rm atm}^2$ in the neutrino sector. 
We note that a small but non-zero type-I contribution is required in the
neutrino mass matrix given by Eq.~(\ref{eq:neu}), in order to have a consistent
fit with the correct mass squared ratio $\Delta m_\odot^2/\Delta m_{\rm
atm}^2$; on the other hand, a large type-I contribution will spoil the TBM
structure given by Eq.~(\ref{eq:f}) and hence results in too small mixing
angles. A balancing of the two is required in order to satisfy the observed
neutrino oscillation data. We also note that SUSY threshold corrections to 
the down quark mass matrix~\cite{bagger}
must be included in order to get a consistent fit for the charged fermion 
sector. The details of this analysis are given in Section V. Also, as discussed 
in Section VI, the proton decay constraints are satisfied only for low $\tan\beta$ in this model, as the Yukawa couplings responsible for the proton decay rates 
grow with $\tan\beta$. Hence, we have chosen $\tan\beta=10$ for our numerical 
analysis given below.
\begin{table}[!ht]
\begin{center}
\begin{tabular}{||c|c||}	\hline\hline
	$M$ (GeV) & 83.06\\
	$m_0$ (GeV) & 1.201 - 0.9007$i$\\
	$m_1$ (GeV) & 0.2033 - 0.01170$i$\\
	$\delta$ (GeV) & 0.2129 + 0.08201$i$\\
	$r_1/ \tan\beta$ & 0.01624\\
	$r_2$ & -0.1382\\
	$r_3$ & 0.1358\\
        $\alpha$ & $5.0^\circ$\\
	\hline\hline
\end{tabular}
\caption{Best fit values for the model parameters at the GUT scale.}
\label{table:params}
\end{center}
\end{table}
\begin{table}[!ht]
\begin{center}
\begin{tabular}{||c|c|c||}\hline\hline
	& best fit & exp value\\
	\hline\hline
	$m_e$ (MeV) & 0.3585 & $0.3585^{+0.0003}_{-0.003}$\\
	$m_\mu$ (MeV) & 75.6717 & $75.6715^{+0.0578}_{-0.0501}$\\
	$m_\tau$ (GeV) & 1.2922 & $1.2922^{+0.0013}_{-0.0012}$\\
	$m_d$ (MeV) & 2.0034 & $1.5036^{+0.4235}_{-0.2304}$\\
	$m_s$ (MeV) & 23.4494 & $29.9454^{+4.3001}_{-4.5444}$\\ 
	$m_b$ (GeV) & 1.0335 & $1.0636^{+0.1414}_{-0.0865}$\\
	$m_u$ (MeV) & 0.8192 & $0.7238^{+0.1365}_{-0.1467}$\\ 
	$m_c$ (MeV) & 207.4990 & $210.3273^{+19.0036}_{-21.2264}$\\
	$m_t$ (GeV) & 82.8964 & $82.4333^{+30.2676}_{-14.7686}$\\
	$V_{us}$ & 0.2245 & $0.2243\pm 0.0016$\\
	$V_{ub}$ & 0.0034 & $0.0032\pm 0.0005$\\
	$V_{cb}$ & 0.0351 & $0.0351\pm 0.0013$\\
	$J$ & $2.052\times 10^{-5}$ & $(2.2\pm 0.6) \times 10^{-5}$\\
	$\Delta m_{\odot}^2 / \Delta m_{\rm atm}^2$ & 0.0311 & $0.0320\pm 0.0025$\\
	\hline
	$\chi^2$ & 3.39 & \\
	\hline\hline
\end{tabular}
\caption{Best fit values for the charged fermion masses and the 
most relevant quark mixing parameters, as well as the 
solar-to-atmospheric mass squared ratio. The $1\sigma$ experimental values~
\cite{das}, with masses and mixings extrapolated by MSSM renormalization group 
(RG) equations to the GUT scale, are also shown for comparison. 
Note that the values of the bottom quark mass and the CKM mixing parameters involving the 
third generation quoted here include the SUSY-threshold corrections (see Section V).} 
\label{table:fit}
\end{center}
\end{table}

The fit results are displayed in Tables \ref{table:params} and \ref{table:fit};
Table \ref{table:params} gives the numerical values of the model parameters
yielding the best fit values shown in Table \ref{table:fit}. Here, $\alpha$ is
the mixing angle for the third generation matter fermion $\psi$ with the
vector-like field $\psi_V$ specific to the model~\cite{DMM}; the limit
$\alpha=0$ gives the form for the mass matrices dictated by $S_4$ symmetry, as
given by Eqs.~(\ref{eq:f}) exactly, and the fit value of $\alpha = 5^\circ$
approximates this limit. With this in mind, note that the top quark mass in the
model is given by 
\begin{eqnarray}
	m_t \simeq U_{uH}\kappa_u h_{33} \cos^2\alpha
	\label{eq:top}
\end{eqnarray}

In the neutrino sector, as noted earlier, the correct mass squared ratio 
$\Delta m_\odot^2/\Delta m_{\rm atm}^2$ as well as large solar and atmospheric 
mixing angles fix the relative size between the type-I and type-II 
contributions, and then the overall scale is determined from the largest mass 
eigenvalue, assuming a normal hierarchy. We find that for the  
best fit parameters shown in Table I, and for $v_R = 2.0\times 10^{16}$ GeV 
(same as the GUT scale), $v_L = 6.810$ eV yields the right neutrino mass 
scale with $m_3\sim 0.05$ eV.
For estimating the proton decay rates as well as for the neutrino masses and 
mixing, we must extract the magnitudes of the 
raw yukawa couplings $h,f,h'$, which can be done using the expressions in 
Eq.~(\ref{eq:couplings}). However, these couplings depend on the vev mixing 
parameters $U_{q\alpha}$, and hence, there is some freedom in their 
determination, although the unitarity constraints on the $U$'s,  
$\sum_\alpha |U_{q\alpha}|^2 \leq 1$, and the top-quark mass relation in this 
model, given by Eq.~(\ref{eq:top}), provide some restriction on these mixing 
parameters. The values chosen for the up-type mixings are 
$U_{uH} = 0.40, U_{u\Delta} = 0.4033,$ and $U_{uH'} = 0.72$, and using the fit 
values for the $r$'s from Table \ref{table:params} and Eq.~(\ref{eq:r}), the resulting values 
for the down-type mixings are $U_{dH} = 0.06497, U_{d\Delta} = -0.4739,$ and 
$U_{dH'} = 0.8611$. Given these values and the running vevs $\kappa_u = 123.8,~
\kappa_d = 17.9$ GeV at GUT-scale~\cite{das}, the resulting dimensionless couplings are found to be
\begin{eqnarray}\label{hfh}
	h &=& \left(\begin{array}{ccc}
		0 & &\\
		 & 0 & \\
		 & & 1.677
	 \end{array}\right) \nonumber \\
f &=& \left(\begin{array}{ccc}
0 & (-5.628+0.3238i)\times 10^{-4} & (-5.628+0.3238i)\times 10^{-4} \\
(-5.628+0.3238i)\times 10^{-4} & (-3.326+2.494i)\times 10^{-3} & (2.763-2.461i)\times 10^{-3} \\
(-5.628+0.3238i)\times 10^{-4} & (2.763-2.461i)\times 10^{-3} & (-3.326+2.494i)\times 10^{-3} \end{array} \right) \nonumber \\
h^\prime &=& \left(\begin{array}{ccc}
0 & (3.243+1.250i)\times 10^{-4} & (-3.243-1.250i)\times 10^{-4}\\
(3.243+1.250i)\times 10^{-4} & 0 & 0 \\
(-3.243-1.250i)\times 10^{-4} & 0 & 0 \end{array} \right) 
\end{eqnarray}

The predicted values for the neutrino mixing parameters corresponding
to the best fit parameter values in the model given in Table \ref{table:params} are summarized in Table \ref{table:neu_fit}.
We find that a consistent fermion sector fit in
this model predicts the reactor mixing angle $\theta_{13}$ to be
non-zero and within a very narrow range $4.5^\circ-5.5^\circ$, which
 is well within the $3\sigma$ lower bound of many
recent experimental results~\cite{T2K,MINOS,dchooz}, but is
only marginally consistent with the latest result from Daya
Bay~\cite{DayaBay} and RENO~\cite{reno}. We show in Section IV that a large 
$\theta_{13}$ value consistent with the Daya Bay and RENO results can be 
obtained in this model with a slightly different vacuum alignment than that 
given in Ref.~\cite{DMM}. 

\begin{table}[h!]
	\begin{center}
	\begin{tabular}{||c|c|c||}\hline\hline
	& predicted value & $3\sigma$ exp range\\
	\hline\hline
	$\theta_{12}$ & $32.34^\circ$ & $(30.6 - 36.8)^\circ$\\ \hline
	$\theta_{23}$ & $49.41^\circ$ & $(35.7 - 53.1)^\circ$\\ \hline
	$\theta_{13}$ & $5.13^\circ$ & $(1.8 - 12.1)^\circ$\\ 
	& & [$(5.9 - 11.6)^\circ$]\\ \hline
	$\delta_{\rm D}$ & $144.4^\circ$ & \\
\hline\hline
\end{tabular}
\end{center}
\caption{The model predictions for the neutrino mixing angles for the 
best fit parameter values given in Table I. We also show the $3\sigma$
range of values from the updated global neutrino data
analysis~\cite{lisinew}, and for $\theta_{13}$, we show in square
brackets the most recent Daya Bay result~\cite{DayaBay}. Note that the predicted
value is only marginally consistent with the $3\sigma$ value of the Daya Bay 
result.}
\label{table:neu_fit}
\end{table}
\section{An Improved Fit with a Different Vacuum Alignment}
In this section, we discuss a different flavon vacuum alignment than that 
presented earlier [cf. Eq.~(\ref{eq:flavev})]. This requires us to choose a specific value of $n$ for the 
$Z_n$ symmetry of the superpotential given by Eq.~(\ref{eq:supmin}). With this 
assumption, we can add the $S_4$-singlet part of 
a linear term like $\phi_2\phi_3$ to the superpotential 
in Eq.~(\ref{eq:supmin}) which upon minimization results in the following 
vacuum structure: 
\begin{eqnarray}
	\phi_1 = \left(\begin{array}{c} 0 \\ 0 \\ 1 
	\end{array}\right),~ 
	\phi_2 = \left(\begin{array}{c} \epsilon \\ a \\ b
	\end{array}\right),~ 
	\phi_3 = \left(\begin{array}{c} x \\ y \\ z
	\end{array}\right).
	\label{eq:flavev2}
\end{eqnarray}
One set of values for the components
given above are $(\epsilon,a,b) = (-0.080,-0.752,0.692)$, and $(x,y,z) = (0.937, 0.928, 0.936)$. Given this shifted flavon vacuum alignment, our mass matrix couplings in
comparison to Eqs.~(\ref{eq:f}) and (\ref{eq:hp}) become
\begin{eqnarray}
\bar{f} &=& m_0 \left(\begin{array}{ccc} \epsilon^2 & \epsilon a & \epsilon b \\
a \epsilon & a^2 & a b \\ b \epsilon & b a & b^2 \end{array}\right) +
m_1 \left(\begin{array}{ccc} 0 & z & y \\ z & 0 & x \\
y & x & 0 \end{array}\right), \label{eq:new_f} \\
\bar{h}' &=& \delta \left(\begin{array}{ccc} 0 & b & a \\
b & 0 & \epsilon \\ a & \epsilon & 0 \end{array}\right),
\label{eq:new_hp}
\end{eqnarray}
with no change to the $h$ coupling.  Performing the $\chi^2$-minimization again
with these new couplings gives a fit with no substantial changes in the
charged sector, but with important improvements to the neutrino sector
predictions. The resulting parameter values for this fit are given in
Table~\ref{table:new_params}, and the best fit values for the masses and
mixings are given in Table~\ref{table:new_fit}.

\begin{table}[!ht]
\begin{center}
\begin{tabular}{||c|c||}	\hline\hline
	$M$ (GeV) & 84.33\\
	$m_0$ (GeV) & 2.607 - 0.3277$i$\\
	$m_1$ (GeV) & -0.3052 - 0.03412$i$\\
	$\delta$ (GeV) & -0.1937 - 0.2719$i$\\
	$r_1/ \tan\beta$ & 0.01591\\
	$r_2$ & -0.1388\\
	$r_3$ & 0.05867\\
        $\alpha$ & $18.5^\circ$\\
	\hline\hline
\end{tabular}
\caption{The improved best fit values for the model parameters at the GUT scale.}
\label{table:new_params}
\end{center}
\end{table}

\begin{table}[!ht]
\begin{center}
\begin{tabular}{||c|c|c||}\hline\hline
	& best fit & exp value\\
	\hline\hline
	$m_e$ (MeV) & 0.3585 & $0.3585^{+0.0003}_{-0.003}$\\
	$m_\mu$ (MeV) & 75.6719 & $75.6715^{+0.0578}_{-0.0501}$\\
	$m_\tau$ (GeV) & 1.2922 & $1.2922^{+0.0013}_{-0.0012}$\\
	$m_d$ (MeV) & 0.8960 & $1.5036^{+0.4235}_{-0.2304}$\\
	$m_s$ (MeV) & 21.9535 & $29.9454^{+4.3001}_{-4.5444}$\\ 
	$m_b$ (GeV) & 1.0627 & $1.0636^{+0.1414}_{-0.0865}$\\
	$m_u$ (MeV) & 0.7284 & $0.7238^{+0.1365}_{-0.1467}$\\ 
	$m_c$ (MeV) & 209.8979 & $210.3273^{+19.0036}_{-21.2264}$\\
	$m_t$ (GeV) & 84.1739 & $82.4333^{+30.2676}_{-14.7686}$\\
	$V_{us}$ & 0.2243 & $0.2243\pm 0.0016$\\
	$V_{ub}$ & 0.0033 & $0.0032\pm 0.0005$\\
	$V_{cb}$ & 0.0351 & $0.0351\pm 0.0013$\\
	$J$ & $2.19\times 10^{-5}$ & $(2.2\pm 0.6) \times 10^{-5}$\\
	$\Delta m_{\odot}^2 / \Delta m_{\rm atm}^2$ & 0.0321 & $0.0320\pm 0.0025$\\
	\hline
	$\chi^2$ & 4.05 & \\
	\hline\hline
\end{tabular}
\caption{The improved best fit values for the charged fermion masses and the 
most relevant quark mixing parameters, as well as the 
solar-to-atmospheric mass squared ratio.
} 
\label{table:new_fit}
\end{center}
\end{table}

The value $v_R$ for this fit is $0.7 \times 10^{16}$ GeV, the value
of $v_L$ was taken as 8.921 eV, and the values for the up-type Higgs
mixings were chosen to be $U_{uH} = 0.35, U_{u\Delta} = 0.63,$ and
$U_{uH'} = 0.25$; given the fit values for the $r$'s from Table
\ref{table:new_params}, the resulting values for the down-type mixings
are $U_{dH} = 0.05568, U_{d\Delta} = -0.7223,$ and $U_{dH'} = 0.6779$.
Using these values and the same prescription as in the previous
section but with the new $f$ and $h^\prime$ coupling definitions, the
resulting dimensionless couplings are now found to be
\begin{eqnarray}\label{new_hfh}
	h &=& \left(\begin{array}{ccc}
		0 & &\\
		 & 0 & \\
		 & & 1.946
	 \end{array}\right) \\
f &=& \left(\begin{array}{ccc}
  (-3.067+0.3855i)\times 10^{-5} & (2.245+0.9247i)\times 10^{-4} &
  (7.649+0.2352i)\times 10^{-4}\\
  (2.245+0.9247i)\times 10^{-4} & (-2.623+0.3297i)\times 10^{-3} &
  (2.923-0.2465i)\times 10^{-3} \\
(7.649+0.2352i)\times 10^{-4} & (2.923-0.2465i)\times 10^{-3} &
(-2.221+0.2792i)\times 10^{-3} \end{array} \right) \nonumber \\
h^\prime &=& \left(\begin{array}{ccc}
0 & (2.541-3.567i)\times 10^{-4} & (-2.762+3.877i)\times 10^{-4}\\
(2.541-3.567i)\times 10^{-4} & 0 & (-2.986+4.192i)\times 10^{-5} \\
(-2.762+3.877i)\times 10^{-4} & (-2.986+4.192i)\times 10^{-5} & 0
\end{array} \right) \nonumber 
\end{eqnarray}

\begin{figure}[h!]
	\includegraphics[width=7.5cm]{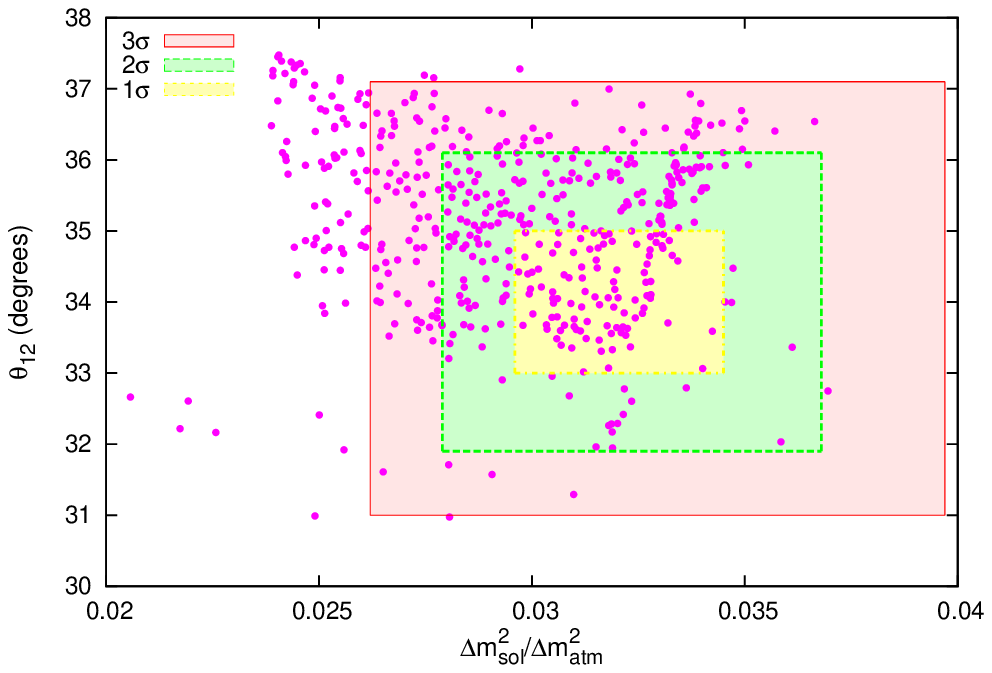}
	\hspace{0.5cm}
	\includegraphics[width=7.5cm]{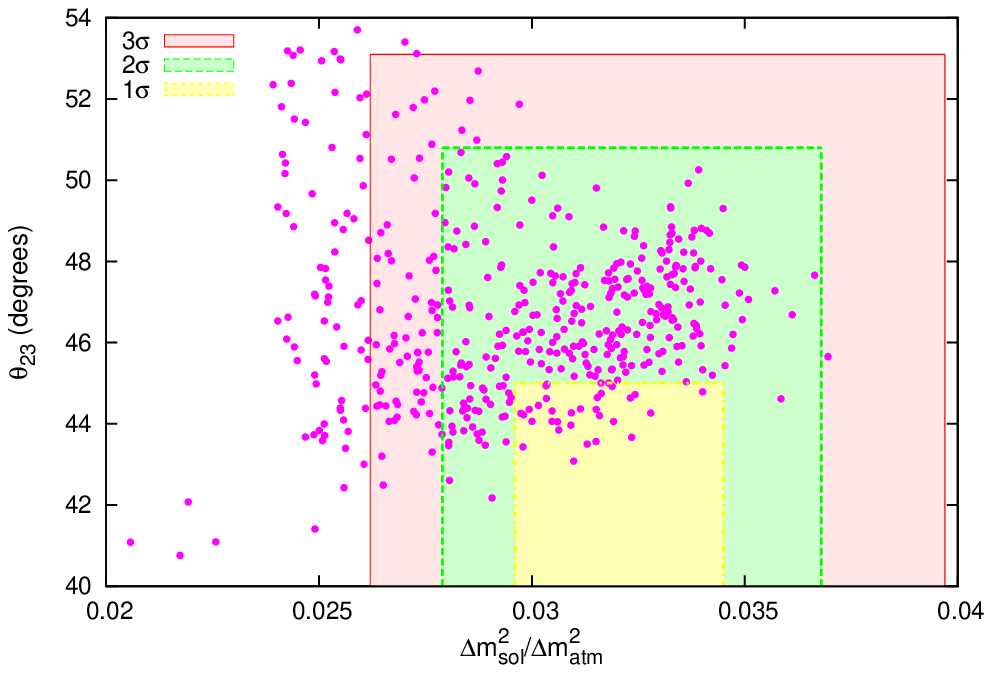}\\
	\includegraphics[width=7.5cm]{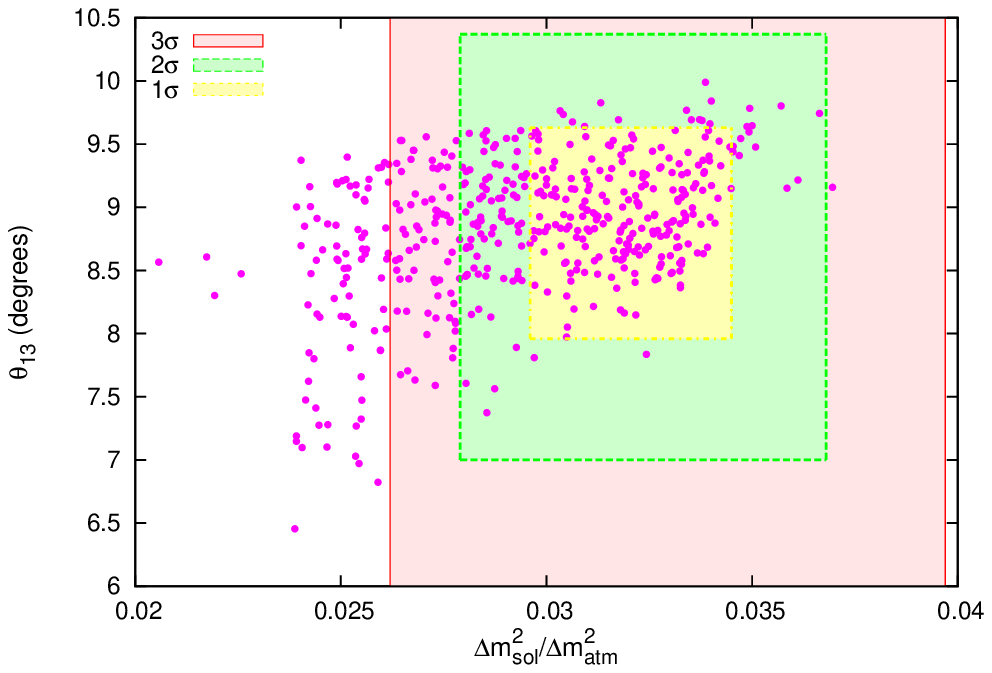}
	\hspace{0.5cm}
	\includegraphics[width=7.5cm]{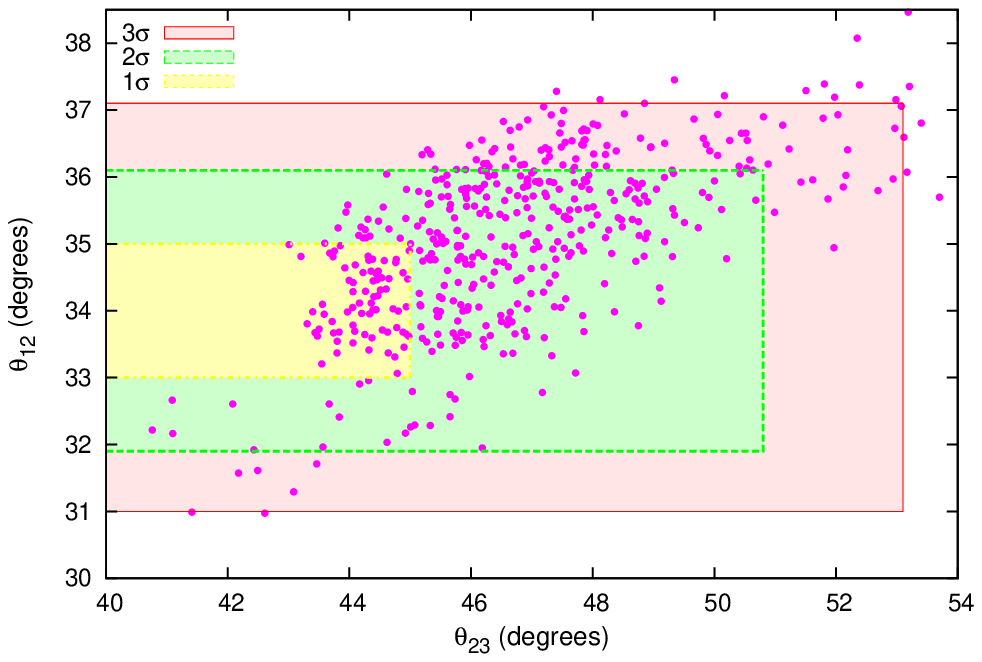}
        \caption{The predicted correlation between the neutrino sector 
	observables, namely $\Delta m_{\odot}^2 / \Delta m_{\rm atm}^2$ and 
	the mixing angles $\theta_{12},\theta_{23},
	\theta_{13}$, based on the fermion sector fit in the model. 
	The $1,2$ and $3\sigma$ experimental limits for $\theta_{12},
	\theta_{23}$ and $\Delta m^2_{\odot}/\Delta m^2_{\rm atm}$ 
	are also shown (shaded regions).}
	\label{fig:thetas}
\end{figure}

The predicted values for the neutrino mixing parameters corresponding to the
best fit parameter values in the model given in Table \ref{table:new_params} are
summarized in Table \ref{table:new_neu_fit}; the correlations between the different parameters in the neutrino sector while satisfying the charged fermion 
constraints are shown in the scatter plots of Figure~\ref{fig:thetas}. 
Notice that in addition to small improvements to the
predicted values for $\theta_{12}$ and $\theta_{23}$ compared to those given 
in Table~\ref{table:neu_fit}, the value for $\theta_{13}$ is now larger 
and consistent within $1\sigma$ of the Daya Bay central value of~$\sim 8.8^\circ$~\cite{DayaBay}.
\begin{table}[h!]
	\begin{center}
	\begin{tabular}{||c|c|c||}\hline\hline
	& predicted value & $3\sigma$ exp range\\
	\hline\hline
	$\theta_{12}$ & $33.77^\circ$ & $(30.6 - 36.8)^\circ$\\ \hline
	$\theta_{23}$ & $44.82^\circ$ & $(35.7 - 53.1)^\circ$\\ \hline
	$\theta_{13}$ & $9.02^\circ$ & $(1.8 - 12.1)^\circ$\\ 
	& & [$(5.9 - 11.6)^\circ$]\\ \hline
	$\delta_{\rm D}$ & $-165.28^\circ$ & \\
\hline\hline
\end{tabular}
\end{center}
\caption{The model predictions for the neutrino mixing angles for the 
best fit parameter values given in Table I. We also show the $3\sigma$
range of values from the updated global neutrino data
analysis~\cite{lisinew}, and for $\theta_{13}$, we show in square
brackets the most recent Daya Bay result~\cite{DayaBay}. Note that the 
predicted value now is consistent with the Daya Bay result.}
\label{table:new_neu_fit}
\end{table}

\section{Threshold Correction and Low Energy Mass Spectrum}
In order to compare the fermion masses and mixing values obtained 
from the model at the GUT scale with the experimental values at the weak scale, 
we must take into account the SUSY threshold correction effects~\cite{poko,bagger}. 
There are two main contributions to the SUSY threshold 
correction to the fermion masses: one coming from the gluino loop and another 
from the chargino loop. The largest correction is for the bottom mass, which is 
given by~\cite{poko}
\begin{eqnarray}
	&&\frac{\delta m_{b}}{m_b} \simeq \epsilon_1 + \epsilon_2|V_{tb}|^2,\\
	&{\rm where~ ~ ~ ~}& 
	\epsilon_1 = \frac{2\alpha_s}{3\pi}\mu m_{\tilde{g}}\tan\beta~ I_3
	(m^2_{\tilde{g}},m_{{\tilde b}_1}^2,m_{ {\tilde b}_2}^2), 
	\label{eq:eps1}\\ 
	&&\epsilon_2 = \frac{1}{16\pi^2}\mu A_t y_t^2 \tan\beta~ I_3(\mu^2,
	m^2_{ {\tilde t}_1}, m^2_{ {\tilde t}_2})
	\label{eq:eps2}
\end{eqnarray}
and the function $I_3$ is given by
\begin{eqnarray}
	I_3(a,b,c) = \frac{ab\log\left(\frac{a}{b}\right) + bc \log\left(\frac{b}{c}\right) + ca \log\left(\frac{c}{a}\right)}
	{(a-b)(b-c)(a-c)}
\end{eqnarray}
Similarly, if we do not add any off-diagonal threshold corrections,  
the CKM mixings involving the third generation receive corrections as follows~
\cite{poko}:
\begin{eqnarray}
	\frac{\delta V_{ub}}{V_{ub}} \simeq 
\frac{\delta V_{cb}}{V_{cb}} \simeq 
\frac{\delta V_{td}}{V_{td}} \simeq 
\frac{\delta V_{ts}}{V_{ts}} \approx -\epsilon_2 
\end{eqnarray}
However, once off diagonal threshold corrections are included, there are further changes to CKM mixing, which we take into account in our 
numerical analysis.

From the numerical fit, we find that at the GUT scale, without including 
the threshold corrections, some of the best-fit values 
predicted by the model do not agree with experimental 
values extrapolated to the GUT scale (see Table II). In particular, 
we find  
\begin{eqnarray}
	m_b = 1.37~{\rm GeV},~ ~|V_{ub}| = 0.0015,~ 
	|V_{cb}| = 0.0160,~|V_{td}| = 0.0047,~|V_{ts}| = 0.0153.
	\label{eq:mbgut}
\end{eqnarray}
Comparing these values with the experimental values, we note that large 
negative threshold corrections are required for the model to have a consistent 
fermion-sector fit. We parametrize the SUSY threshold corrections at the GUT 
scale by modifying the third generation elements of the 
down-quark mass matrix as follows:
\begin{eqnarray}
	{\cal M}'_d = {\cal M}_d+\delta{\cal M}_d, ~{\rm where}~ ~
	\delta{\cal M}_d = \frac{r_1}{\tan\beta}\left(\begin{array}{ccc}
		0 & 0 & \delta_{13}\\
		0 & 0 & \delta_{23}\\
		\delta_{13} & \delta_{23} & \delta_{33}
	\end{array}\right)
\end{eqnarray}
and ${\cal M}_d$ is given by Eq.~(\ref{eq:mass}). The required threshold 
corrections at the GUT scale are: 
\begin{eqnarray}
	\delta_{13} = 0.09~{\rm GeV}, ~\delta_{23} = -0.96~{\rm GeV},~
	\delta_{33} = -20.68~{\rm GeV}.
	\label{eq:thresgut}
\end{eqnarray}
Note that these threshold corrections, when extrapolated down to the weak scale and added to the RG-extrapolated values of the $b$-quark mass and the CKM 
mixing parameters, yield results within $1\sigma$ range of the experimental values at $M_Z$.   

At the weak scale, it is clear from Eqs.~(\ref{eq:eps1}) and (\ref{eq:eps2}) 
that for the large negative threshold corrections given by Eq.~(\ref{eq:thresgut}), we must have $\mu<0$ if the gluino 
term is dominant, or opposite signs for 
$\mu$ and $A_t$ if the chargino contribution is dominant.  
These observations have important consequences for the MSSM light 
Higgs mass as well as the sparticle spectrum, as shown below.

The one-loop radiative correction to the MSSM light Higgs mass is given by
~\cite{haber}
\begin{eqnarray}
	m_h^2 \simeq m_Z^2\cos^2{2\beta}+\frac{3g_2^2m_t^4}{16\pi^2m_W^2}
	\log{\left[
	\left(1+\frac{m_{ {\tilde t}_1}^2}{m_t^2}\right)
	\left(1+\frac{m^2_{ {\tilde t}_2}}{m_t^2}\right)\right]}
\end{eqnarray}
where $m_{\tilde t_{1,2}}$ are the stop mass eigenvalues, which are obtained 
by diagonalizing the matrix 
\begin{eqnarray}
	{\cal M}^2_{\tilde t} = \left(\begin{array}{cc}
		m^2_{\tilde t_{L}}+ m_t^2 + m_Z^2\cos{2\beta}\left(\frac{1}{2}-\frac{2}{3}
	\sin^2\theta_W\right), & 
		m_t(-A_t+\mu \cot{\beta}) \\
		m_t(-A_t+\mu \cot{\beta}) & 
		m^2_{\tilde t_{R}}+ m_t^2 +  m_Z^2\cos{2\beta}\left(\frac{2}{3}
	\sin^2\theta_W\right)
	\end{array}\right)\nonumber
\end{eqnarray}

\begin{figure}[h!]
	\centering
	\includegraphics[width=8cm]{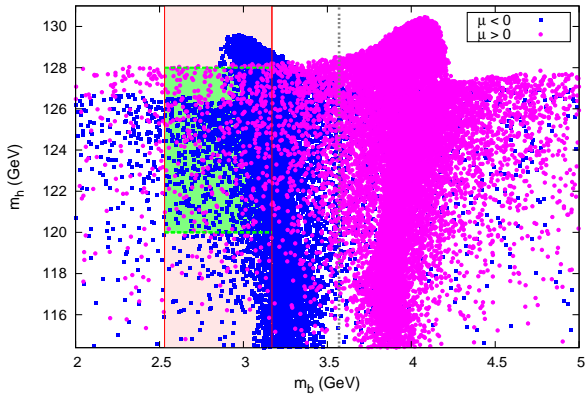}
	\caption{Higgs mass prediction in the model required by the bottom 
	quark mass fit, including the threshold correction effects. 
	The dashed vertical line is the best fit 
	value for $m_b (M_Z)$ without the threshold corrections.
	The vertical (red) shaded region is the $1\sigma$ 
	experimental range of $m_b (M_Z) = (2.85\pm 0.32)$ GeV, obtained from 
	the LEP data~\cite{LEP}. The horizontal (green) shaded region shows the 
	range in which $\sim 3\sigma$ excess of events for SM-like Higgs 
	were observed recently at the LHC~\cite{lhchiggs}.}
	\label{fig:hb}
\end{figure}

Note that the same mass parameters, namely the gluino and third generation 
squark masses as well as the trilinear term $A_t$, appear in the threshold correction and 
Higgs mass correction. Given that this model requires a large threshold 
correction to have a consistent fit in the fermion sector, 
we expect some correlation between the two corrections. To show it 
quantitatively, we have chosen the simplest case of 
mSUGRA-type GUT scale spectrum, as an illustration, although our results do not depend 
on the assumption of mSUGRA.
Here we scan over the parameter space for 
$200~{\rm GeV}\leq m_0 \leq 2~{\rm TeV},~200~{\rm GeV}\leq m_{1/2} \leq 
2~{\rm TeV}$, $-5~{\rm TeV} \leq A_0 \leq 5~{\rm TeV}$ and $\tan\beta=10$, 
using the {\tt ISAJET} package~\cite{isajet}. 
The results are shown in Fig.~\ref{fig:hb}, from which we find that the 
lightest 
Higgs mass is required to be below 129 (128) GeV for $\mu<0 ~(>0)$ in order 
to have the right amount of threshold correction to satisfy 
the bottom-quark 
mass constraint (vertical red shaded region) in this model. The horizontal green shaded region 
shows the range of Higgs mass in which a mild excess of events has been recently 
reported at the LHC~\cite{lhchiggs}.   

\begin{figure}[h!]
	\centering
	\includegraphics[width=7.5cm]{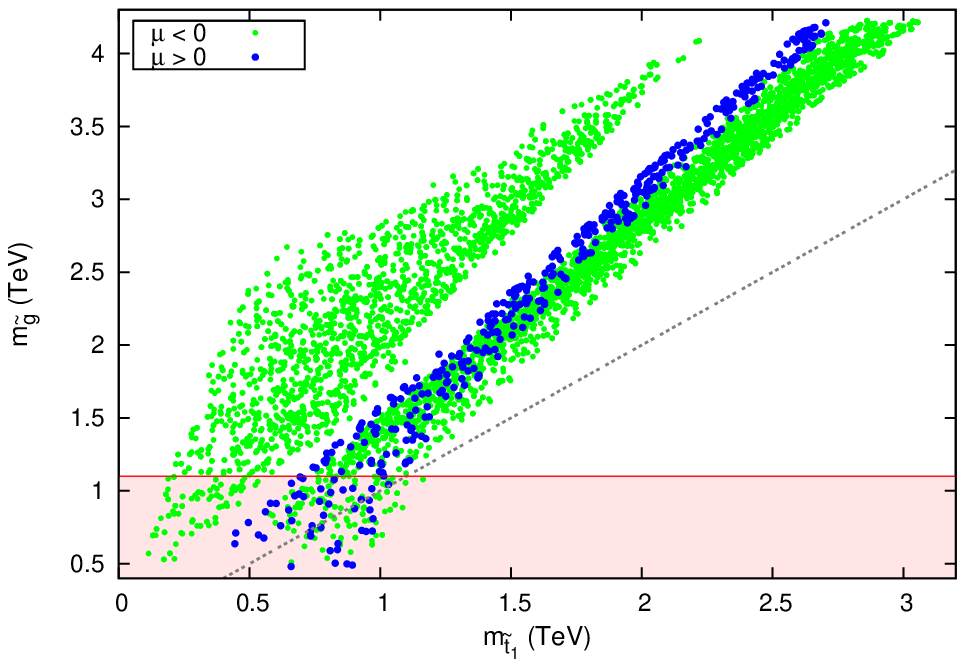}
	\hspace{0.5cm}
	\includegraphics[width=7.5cm]{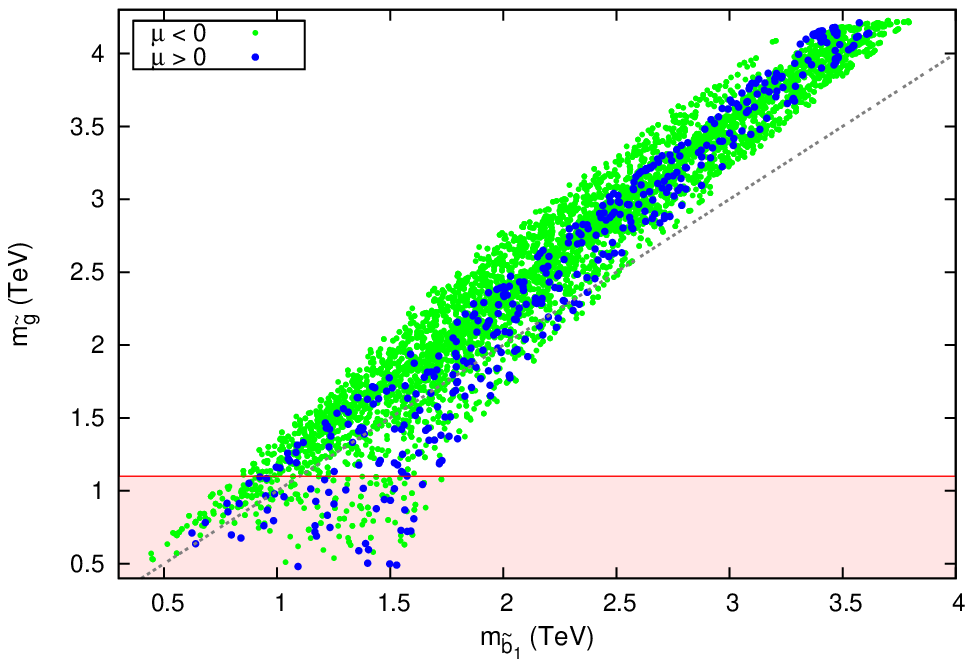}
	\caption{The correlation between the gluino mass and the light 
	stop and sbottom masses, 
	corresponding to the shaded region in Fig.~\ref{fig:hb}.
The dashed line is for gluino mass equal to the relevant squark mass in the plot. 
The red shaded region is ruled out by the current LHC data~\cite{cms}.}
	\label{fig:stb}
\end{figure}

The correlations between the gluino mass and the light stop and sbottom masses 
for the required threshold corrections (shaded region in Fig.~\ref{fig:hb}) 
are shown in Fig.~\ref{fig:stb}. We 
find that the large threshold correction requirement in this model requires 
the gluino to be always heavier than the light stop, but not necessarily 
heavier than the light sbottom. Moreover, for gluino masses satisfying the 
current LHC lower bound of 1.1 TeV~\cite{cms} and for Higgs mass between 120-128 GeV, 
we find a lower limit for the stop mass of 755 (211) GeV for $\mu>0 (<0)$ and 
similarly for the sbottom mass of 1013 (895) GeV. The milder limit on the 
squark masses for $\mu<0$ is because of the fact that in this case, the required 
negative threshold corrections can be obtained from both gluino and chargino contributions 
(c.f. Eq.~(\ref{eq:eps1}) and (\ref{eq:eps2})), thus allowing for the $A_t, A_b$ 
values necessary for light stop and sbottom masses, respectively. 
However, for $\mu>0$ case, the gluino contribution is of the wrong sign, and hence we must have very large negative $A_t$ values to obtain the required 
threshold corrections.     

\section{Expectations for proton decay}
We now turn to a discussion of proton lifetime in our model.  As experimental limits on proton lifetimes keep increasing, many simple SUSY-GUT models have 
either been ruled out or become more and more constrained. It is therefore important to ensure that any GUT model for neutrino masses is consistent with those limits. As is well known, the dominant contributions to proton decay
comes from color triplet Higgsino exchange in these modes~\cite{review} and can in general lead to large proton decay amplitudes~\cite{babu2}.  It was suggested in Ref.~\cite{dmm1} that
one way to suppress this amplitude without invoking cancellations is to choose appropriate flavor structure for the Yukawa couplings. The current model falls into this category where the existence of small elements in $h$, $h^\prime$ and $f$ matrices leads to the expectation that we should be able to satisfy the proton decay constraints without any cancellation. Let us now see how this occurs using the fit for Yukawa couplings we obtained in the previous section.

 The  colored Higgs triplets, $\phi_T+\phi_{\bar{T}}$: $((3,1,-1/3)+{\rm c.c.})$ responsible for proton decay in our model arise from 
 ${\bf 10}+{\bf 10}^\prime+{\bf 126}+{\overline{\bf 126}}+{\bf 210}$ multiplets. Once  the triplet fields $\phi_{T}$ and $\phi_{\bar{T}}$ are integrated out,  both $LLLL~(C_L)$ and $RRRR~(C_R)$ operators leading to proton decay emerge :
\begin{equation}
	W_5={1\over{2}}C_L^{ijkl}\ell_kq_lq_iq_j+C_R^{ijkl}e^c_ku^c_lu^c_id^c_j
\end{equation}
The color triplet  $\phi_{T,{\bar{T}}}$ fields are linear combinations of six fields, two of them arising from two ${\bf 10}$'s, three of them arising  from ${\bf 126+{\overline {\bf 126}}}$ and one from {\bf 210}. This leads to a 
$6\times 6$ dimensional mass matrix for the triplets: 
$(\phi_{\bar{T}})_a(M_T)_{ab}(\phi_T)_b$. One can write the dimension five operators in terms of the couplings $h$, $h'$ and $f$ as follows: 
\begin{eqnarray}
	C_L^{ijkl}&=& c h_{ij}h_{kl}+x_1h_{ij}h^\prime_{kl}+x_2 h^\prime_{ij}h_{kl}+x_3 h^\prime_{ij}h^\prime_{kl}+x_4f_{ij}f_{kl}
	\nonumber\\
	&&  ~~ 
	+ x_5f_{ij}h_{kl} +x_6 h_{ij}f_{kl}+x_7h^\prime_{ij}f_{kl}+x_8f_{ij}h^\prime_{kl}.
	\label{eq:chf}
\end{eqnarray} 
Similarly, we can write $C_{RRRR}$ operator 
(and change $x_i$'s to different coefficients $y_i$). The coefficient $c$ is $(M^{-1}_T)_{11}$ and the coefficients $x_i$ and $y_i$ are also given by the components of $M^{-1}_T$. The proton decay amplitude can be written as 
\begin{equation}
	A=\frac{\alpha_2\beta_p\tilde{A}}{4\pi M_T}\frac{m_{\tilde{W}}}
	{m_{\tilde q}^2},
\end{equation}
where ${\tilde A}=c{\tilde A_{hh}}+x_1 {\tilde A_{hf}}+...$ using the c's and $x_i$s given in Eq.~(\ref{eq:chf}), and $\beta_p$ is the nucleon matrix element of the three-quark operator. In our calculation we use $m_{\tilde{q}}=
1.3$ TeV as a typical value for the first two 
generation squark masses (to satisfy the LHC lower bound), and similarly 
$m_{\tilde W}=200$ GeV as a typical wino mass. The proton 
decay width can be written simply as : 
\begin{eqnarray}
	\Gamma \simeq (2.7\times 10^{-50}~{\rm GeV})|C|^2 |f(F,D)|^2
	\left(\frac{2\times 10^{16}~{\rm GeV}}{M_T}\right)^2 \left(\frac{m_{\tilde{W}}}
	{200~{\rm GeV}}\right)^2 \left(\frac{1~ {\rm TeV}}{m_{\tilde{q}}}\right)^4,  
\end{eqnarray}
where $f(F,D)$ are the hadronic form factors, typically of ${\cal O}(1)$. Assuming that the colored Higgs from {\bf 10}'s are the lightest, we find the largest contribution to the $p\rightarrow \bar{\nu}_\mu K^+$ to be arising from $h^{\prime}_{12}$. 
Using the values of $h^\prime$ from the Eq.~(\ref{hfh}) and varying $M_T$ within  a factor 5 of the GUT scale, we find that the partial lifetime of proton in this mode can be as large as $7\times 10^{33}$ years. If however, we lower $\tan\beta$ to 5 ($h^{\prime}_{12}$ will be smaller in this case), or raise the 
squark mass to 1.5 TeV, the lifetime can be be as large as $10^{34}$ years. The partial lifetimes for other flavors of $\nu$ are found to be larger. Similarly, the partial lifetimes for other decay modes, e.g., $n\rightarrow \bar{\nu}_{\mu} \pi^0$ are found to be much larger of ${\cal O}(10^{38})$ years.
We should emphasize here that, in order to generate these numbers, we did not invoke any cancellation. The smallness of the elements of the Yukawa coupling matrices  are sufficient in suppressing the decay rates. We note that the lifetime of the mode $p\rightarrow\bar{\nu}_{\mu} K^+$ is within the search limit of the proposed Hyper-Kamiokande experiment~\cite{abe}.

\section{Summary} We have analyzed the predictions of a minimal
$SO(10)\times S_4$ model of flavor with dominant type II seesaw form
for neutrino masses, where the forms of the Yukawa couplings for the
two {\bf 10} Higgs fields and one {\bf 126} Higgs field are determined
dynamically by flavon vevs at the minimum of the $S_4$-invariant
flavon potential with an additional $Z_n$ symmetry.  The model has
eleven parameters including complex phases and is, thus, a relatively
economical one when compared to other models discussed in the
literature. It gives a very good fit to the charged fermion masses and
the CKM parameters, and it also predicts the neutrino mixing angles
$\theta_{12}$, $\theta_{23}$ as well as $\Delta m^2_\odot/\Delta
m^2_{\rm atm}$ in agreement with observation.  Furthermore, it
predicts a non-zero value for $\theta_{13}$ between
$6^\circ-10^\circ$, which is in the current experimentally preferred
range.  With more accurate determination of $\theta_{13}$ and its
correlation with $\theta_{23}$, the model could be tested in near
future.  The model also predicts a normal hierarchy for the neutrinos
and hence an effective neutrino mass in neutrino-less double beta
decay which is a few milli-electron-volts and is thus not observable
in the current round of the searches for this process. The proton
lifetime for $p\rightarrow{\bar\nu}_\mu k^+$ decay mode can be
$~10^{34}$ years. Finally, the successful fit for fermion masses
require the Higgs mass to be below 129 GeV in this model and put lower
bounds on the third generation squark masses which are well within the
reach of LHC.
\section{Acknowledgment} The  work of P. S. B. D., R. N. M., and M. S.
is supported by the National Science Foundation grant number
PHY-0968854 and the work of B. D. is supported by the DOE grant
DE-FG02-95ER40917. P. S. B. D. and M. S. would like to thank Shabbar
Raza and Mike Richman for their patient help with the numerical tools. 

\end{document}